\documentclass[conference]{IEEEtran}

\usepackage[a4paper, top=1.9cm, bottom=4.4cm, left=2cm, right=2cm]{geometry}

\usepackage{multicol}
\usepackage{amsmath}
\usepackage{graphicx}
\usepackage[numbers]{natbib}

\begin{document}
	
	\title{Fingerprinting Browsers in Encrypted Communications}

	\author{
		\IEEEauthorblockN{
			Sandhya Aneja\IEEEauthorrefmark{1} and
			Nagender Aneja\IEEEauthorrefmark{2}
		}
		
		\IEEEauthorblockA{
			\IEEEauthorrefmark{1}School of Computer Science and Mathematics, Marist College, Poughkeepsie, NY, USA\\
			\IEEEauthorrefmark{2}Bradley Department of Electrical and Computer Engineering, Virginia Tech, Blacksburg, VA, USA
		}
		
		\IEEEauthorblockA{
			E-mail:  sandhya.aneja@marist.edu, naneja@vt.edu}
	}

	\maketitle
	
	\begin{abstract}
		Browser fingerprinting is the identification of a browser through the network traffic captured during communication between the browser and server. This can be done using the HTTP protocol, browser extensions, and other methods. This paper discusses browser fingerprinting using the HTTPS over TLS 1.3 protocol. The study observed that different browsers use a different number of messages to communicate with the server, and the length of messages also varies. To conduct the study, a network was set up using a UTM hypervisor with one virtual machine as the server and another as a VM with a different browser. The communication was captured, and it was found that there was a 30\%-35\% dissimilarity in the behavior of different browsers.
	\end{abstract}
	
	
	\section{Introduction}
	Identifying the client device information is crucial for the server, and this process was made easier by including the user-agent header field in the HTTP protocol \cite{rfc2068}. Browser fingerprinting \cite{zhang2022survey} is common using the user-agent field when a client communicates with a server using HTTP since HTTP communication is in plain text and is so easier to classify the network traffic. The browser uses many APIs, such as canvas, WebGL, web Audio APIs, and browser extensions for user customization, such as a password manager and video downloader. All these attributes can be used to define a browser's uniqueness. Browser fingerprinting is useful because it helps identify not only the client but also malicious users, as their fingerprints are different from those of legitimate users.
	
	This research is for browser fingerprinting over encrypted communication using the HTTPS protocol \cite{rfc2660}. HTTPS protocol uses TLS encryption, which helps authenticate the parties involved in the communication, increasing the reliability and trustworthiness of the connection. However, HTTPS traffic complicates legitimate network monitoring. Moreover, it has become complicated to detect the malware since malware has started using HTTPS. \citet{husak2016https, garn2019browser} has proposed the classification using the TLS handshake. However, the method involves decryption of HTTPS fields by considering all the possible encryption parameters exchanged during the handshake. 
	
	Our approach is based on the idea that the list of encryption algorithms used (cipher suite list) affects how a browser communicates with the server, which in turn affects the message length of the communication. We treat the message length of a page's communication with a browser as a vector, resulting in n vectors for n browsers. We employ interpolation to standardize the lengths of the vectors and calculate the dissimilarity between browsers.
	
	\begin{enumerate}
		\renewcommand{\theenumi}{\alph{enumi}}
		\item We observed the length of TLS handshake and data messages are different for different browsers for a same page.
		
		\item We used interpolation to make the length vector of a browser messages to compare with others 
		
		\item We used cosine similarity to compare the browsers
		
		\item We observed the difference in length is due to cipher suite list used be TLS protocol
	\end{enumerate}
	
	\section{Preliminaries}
	Transport Layer Security \cite{wiki2024tls} protocol extends secure socket layer protocol. Its current version is 1.3. This protocol uses two phases wherein first phase is handshake between client and server to negotiate the encryption and integrity algorithms. The second phase is the sending the application data using the key exchanged in first phase. In the first phase, once integrity and encryption algorithms  are negotiated, the keys are exchanged, the data is exchanged. However, the record protocol parameters still cover the application data.  Encryption, key exchange and integrity algorithms are changed from 1.2 to 1.3 version \cite{wiki2024tls} .
	
	\section{Related Work} \label{relatedwork}
	\citet{husak2016https} assumed that a client uses both type of traffics HTTP and HTTPS. The authors collected the network traffic from the devices on the campus and a installed web server. The authors paired HTTPS and HTTP traffic from an IP address. They correlated the cipher suite list from HTTPS with user-agent from HTTP traffic. By pairing the user-agents with the cipher suite list, they were able to identify the user-agent, host operating system, and user application for all pairs within the HTTPS traffic. In their traffic, there were 307 pairs with 273 different browsers. They could identify the user-agent in the 95\% of traffic with  most commonly used 10 browsers. This method is compute intensive as it requires using all encryption algorithms of the cipher suite list.
		
	\citet{garn2019browser} proposed combinatorial  sequence based browser fingerprinting for encrypted traffic of HTTPS protocol.  The authors used a Sequence covering array, $SCA(N, S, t)$ matrix of dimension $N \times S$ with entries from a finite set $S$ of $s$ symbols. Every $t-$way permutation of symbols  from $S$ occurs in a atleast one row and each row is a permutation of the $s$ symbols. For example, a $3$-event sequences of six events occur in $10$ ways instead of $20$ ways. The authors used $t$-event sequence of cipher suite list of n-algorithms and tested for pairing it with the browser instead of all n! permutations.

	\citet{laperdrix2020browser} presented a survey on research conducted on browser fingerprinting. The paper presents browser fingerprinting using HTTP headers and Application programming interfaces (API) used by the browser. The authors have discussed the advantages of browser fingerprinting that how the server can run a script to get the differences in the environments of the browser to identify user. This survey does not include browser fingerprinting research for encrypted traffic. \\
	
	\section{Experimental Setup}
	We installed a hypervisor UTM \cite{utm2024} on a Apple Mac  machine with M1 processor and installed three virtual machines Windows 11, Kali Linux and Ubuntu using UTM. The hypervisor provides a virtual Local area network between virtual machines with host machine as a switch to connect the virtual machines to the Internet. We installed Apache web server on the Kali Linux with 6 web pages. We installed three browsers on Windows 11 and two on Ubuntu and accessed the web pages on these with simultaneously capturing the communication at Wireshark.  We configured openSSL  also on the Kali Linux to use generate 4096 byte key and certificate for web server to communicate with browsers using the same credentials for a HTTPS communication. The openSSL use TLS 1.3 version protocol to generate key and certificate. All the browsers also use TLS 1.3 protocol thus the communication of web pages was using TLS 1.3 protocol.

	\subsection{Data Preparation}
	We installed Python and TShark on the Ubuntu machine to extract the fields of TLS 1.3 protocol in a csv file and extracted the fields using Python  to compare the behavior of the browsers. We observed that browsers were using different length sized packets for the communication of a same page.  We extracted lengths of all the messages used by browsers for all 6 web pages in the form of a vector. Thus, we were having 6 vectors for each browser. 
	
	\subsection{Dataset and Methodology} The browsers use different size messages and different number of messages to communicate because of TLS protocol for the communication.  Thus the vectors from the browsers for a web page were of different lengths. We used cosine similarity to present the dissimilarity of browsers. We used interpolation function of Python to make the vectors of equal sizes for the cosine similarity presentation.  The Tables \ref{tab:cosine_similarity} and \ref{tab:dissimilarity} represents the cosine similarity and dissimilarity of the browsers respectively.\\
	
	\begin{table}
		\caption{Similarity between different browsers for  URLs}
		\label{tab:cosine_similarity}
	\begin{tabular}{|p{1.4cm}|p{0.7cm}|p{0.7cm}|p{0.7 cm}|p{0.7cm}|p{0.7cm}|p{0.7cm}|}
		\hline
		Browsers  & url-1& url-2& url-3 & url-4 & url-5 & url-6  \\ \hline \hline
		
		Chrome-Edge & 0.480 & 0.771 & 0.957 & 0.427 & 0.999 & 0.601  \\ \hline
		
		Chrome-Firefox & 0.559 & 0.539 & 0.950 & 0.526 & 0.608 & 0.801\\ \hline
		
		Edge-Firefox  & 0.861 & 0.620 & 0.932 & 0.507 & 0.601 & 0.509   \\ \hline
		
	\end{tabular}
	\end{table}

\begin{table}
	\caption{Dissimilarity between different browsers for URLs}
	\label{tab:dissimilarity}
	\begin{tabular}{|p{1.4cm}|p{0.7cm}|p{0.7cm}|p{0.7cm}|p{0.7cm}|p{0.7cm}|p{0.7cm}|}
		\hline
		Browsers  & url-1 & url-2 & url-3 & url-4 & url-5 & url-6 \\ \hline \hline
		
		Chrome-Edge & 0.520 & 0.229 & 0.043 & 0.573 & 0.001 & 0.399 \\ \hline
		
		Chrome-Firefox & 0.441 & 0.461 & 0.050 & 0.474 & 0.392 & 0.199 \\ \hline
		
		Edge-Firefox & 0.139 & 0.380 & 0.068 & 0.493 & 0.399 & 0.491 \\ \hline
	\end{tabular}
\end{table}

	\section{Results and Discussion} \label{implementation}
	Cosine similarity is a metric used to measure how similar two vectors are. It is often used in fields such as information retrieval, text analysis, and machine learning. The cosine similarity is defined as cosine of angle between two vectors in a multi-dimensional space. The resulting value can be from $-1$ to $1$, where the values $1, 0, -11$ indicates vectors are identical, orthogonal (no similarity), and  diametrically opposite respectively. However, in our case, the range of metric is between $0$ and $1$ since the vector formed with length of messages communicated for handshake of TLS protocol is greater than zero.  Thus, the cosine similarity would be between $0$ and $1$. 
	
\begin{align}
	\text{cosine similarity}(\vec{A}, \vec{B}) &= \frac{\vec{A} \cdot \vec{B}}{\|\vec{A}\| \|\vec{B}\|} \notag \\
	\text{cosine dissimilarity}(\vec{A}, \vec{B}) &= 1 - \frac{\vec{A} \cdot \vec{B}}{\|\vec{A}\| \|\vec{B}\|}
\end{align}

Equations \ref{eqn:v1}	and \ref{eqn:v2} represent two example vectors corresponding to different browsers for the same webpage. If necessary, the vectors are interpolated to ensure equal lengths before calculating the cosine similarity.

\begin{align}
	\label{eqn:v1}
	\mathbf{v_1}^T =
	\begin{bmatrix}
		327 & 1514 &  70 & 84 & 327 & 1514 & \\
		70 & 84 & 391 & 1514 & 70 & 84 & \\
		295 & 1514 & 70 & 118 &  146 &  539 & \\ 
		133 & 133 & 104 & 85 & 85 & 350 & \\
		220 & 69 & 122 &  402 & 100 & 78 
	\end{bmatrix}
\end{align}

\begin{align}
		\label{eqn:v2}
	\mathbf{v_2}^T =
	\begin{bmatrix}
		295  & 1514  &  70  & 84  & 359  &  1514  &  \\
		70  & 84  & 359  &  1514  & 70  &  118  &  \\
		146  & 549  & 133  & 183  & 85  & 85  & \\
		350  &  220  & 92  & 122  &  402  &  100  & \\
		78 
	\end{bmatrix}
\end{align}

We observe that for a web page the dissimilarity may be high up to maximum 52\% while it may be low as .01 \%.  The probability of low dissimilarity is  .05 which is also very small. We can see that we get Chrome-Edge mean dissimilarity as 30.94\%, Chrome-Firefox mean  dissimilarity as 33.57\% and  Edge-Firefox mean disimilarity as 32.77\%.  Thus, the proposed approach is able to distinguish the three browsers upto a large extent.
	
\section{Conclusion and Future Work} 
The browser fingerprinting is important not only from the server perspective to recognize the difference in environment of browsers but also it is required to identify the malicious users. Therefore it is required to study browser fingerprinting with encrypted communication also for risk analysis. In future, we propose to extend the work to many existing web browsers and by capturing the traffic over campus and apply machine learning over a large dataset.  Looking into the features from other protocols such as TCP protocol can improve the accuracy of browser fingerprinting.
	
\bibliographystyle{IEEEtranN}
\bibliography{IEEEabrv, rpaper}

\begin{thebibliography}{8}
\providecommand{\natexlab}[1]{#1}
\providecommand{\url}[1]{#1}
\csname url@samestyle\endcsname
\providecommand{\newblock}{\relax}
\providecommand{\bibinfo}[2]{#2}
\providecommand{\BIBentrySTDinterwordspacing}{\spaceskip=0pt\relax}
\providecommand{\BIBentryALTinterwordstretchfactor}{4}
\providecommand{\BIBentryALTinterwordspacing}{\spaceskip=\fontdimen2\font plus
\BIBentryALTinterwordstretchfactor\fontdimen3\font minus
  \fontdimen4\font\relax}
\providecommand{\BIBforeignlanguage}[2]{{%
\expandafter\ifx\csname l@#1\endcsname\relax
\typeout{** WARNING: IEEEtranN.bst: No hyphenation pattern has been}%
\typeout{** loaded for the language `#1'. Using the pattern for}%
\typeout{** the default language instead.}%
\else
\language=\csname l@#1\endcsname
\fi
#2}}
\providecommand{\BIBdecl}{\relax}
\BIBdecl

\bibitem[Fielding et~al.(1997)Fielding, Nielsen, Mogul, Gettys, and
  Berners-Lee]{rfc2068}
\BIBentryALTinterwordspacing
R.~T. Fielding, H.~Nielsen, J.~Mogul, J.~Gettys, and T.~Berners-Lee,
  ``{Hypertext Transfer Protocol -- HTTP/1.1},'' RFC 2068, Jan. 1997. [Online].
  Available: \url{https://www.rfc-editor.org/info/rfc2068}
\BIBentrySTDinterwordspacing

\bibitem[Zhang et~al.(2022)Zhang, Zhang, Bu, Chen, Sun, and
  Wang]{zhang2022survey}
D.~Zhang, J.~Zhang, Y.~Bu, B.~Chen, C.~Sun, and T.~Wang, ``A survey of browser
  fingerprint research and application,'' \emph{Wireless Communications and
  Mobile Computing}, vol. 2022, no.~1, p. 3363335, 2022.

\bibitem[Rescorla and Schiffman(1999)]{rfc2660}
\BIBentryALTinterwordspacing
E.~Rescorla and A.~Schiffman, ``{The Secure HyperText Transfer Protocol},'' RFC
  2660, 1999. [Online]. Available:
  \url{https://datatracker.ietf.org/doc/html/rfc2660}
\BIBentrySTDinterwordspacing

\bibitem[Hus{\'a}k et~al.(2016)Hus{\'a}k, {\v{C}}erm{\'a}k, Jirs{\'\i}k, and
  {\v{C}}eleda]{husak2016https}
M.~Hus{\'a}k, M.~{\v{C}}erm{\'a}k, T.~Jirs{\'\i}k, and P.~{\v{C}}eleda,
  ``{HTTPS traffic analysis and client identification using passive SSL/TLS
  fingerprinting},'' \emph{EURASIP Journal on Information Security}, vol. 2016,
  pp. 1--14, 2016.

\bibitem[Garn et~al.(2019)Garn, Simos, Zauner, Kuhn, and
  Kacker]{garn2019browser}
B.~Garn, D.~E. Simos, S.~Zauner, R.~Kuhn, and R.~Kacker, ``Browser
  fingerprinting using combinatorial sequence testing,'' in \emph{Proceedings
  of the 6th Annual Symposium on Hot Topics in the Science of Security}, 2019,
  pp. 1--9.

\bibitem[Wikipedia()]{wiki2024tls}
Wikipedia, ``Transport layer security.''

\bibitem[Laperdrix et~al.(2020)Laperdrix, Bielova, Baudry, and
  Avoine]{laperdrix2020browser}
P.~Laperdrix, N.~Bielova, B.~Baudry, and G.~Avoine, ``Browser fingerprinting: A
  survey,'' \emph{ACM Transactions on the Web (TWEB)}, vol.~14, no.~2, pp.
  1--33, 2020.

\bibitem[UTM()]{utm2024}
UTM, ``Virtual machines for mac.''

\end{thebibliography}
	
\end{document}